\documentstyle[12pt]{article}
\begin{document}
\begin{titlepage}
\title{{\bf Precision bounds on $m_H$ and $m_t$}\thanks{Work partially
supported by CICYT under contracts AEN90-0139, AEN90-0683.}}

\author{{\bf F. del Aguila}   \\
Departamento de F\'{\i}sica Te\'orica y del Cosmos \\
Universidad de Granada, E-18071 Granada, Spain
\vspace{.5cm} \\
{\bf M. Mart\'{\i}nez} \\
PPE Division, CERN \\
CH-1211 Geneva 23, Switzerland \\ and \vspace{.5cm} \\
{\bf M. Quir\'os}\thanks{Permanent address:
Instituto de Estructura de la Materia, CSIC, Serrano 123, E-28006
Madrid, Spain.}  \\
Theory Division, CERN \\
CH-1211 Geneva 23, Switzerland}

\date{}
\maketitle

\begin{abstract}
We perform a fit to precise electroweak
data to determine the Higgs and
top masses. Penalty functions  taking into account their
production limits are included. We find
${\displaystyle m_H=65^{+245}_{-4}\ GeV}$  and
${\displaystyle m_t=122^{+25}_{-20}\ GeV}$.
However whereas the top $\chi^2$ distribution behaves properly
near the minimum, the Higgs $\chi^2$ distribution does not,
indicating a statistical fluctuation or new physics. In fact
no significative bound on the Higgs mass can be given at present.
However, if the LEP accuracy is improved
and the top is discovered in the preferred range of top
masses, a meaningful bound on
the Higgs mass could be obtained within the standard model framework.
\end{abstract}

\vspace{3cm}
\noindent
CERN-TH.6389/92 \\
IEM-TH-52/92 \\
February 1992

\vskip-22.5cm
\rightline{CERN-TH.6389/92}
\rightline{IEM-TH-52/92}

\end{titlepage}

\section*{Introduction}

\noindent
The standard model of electroweak interactions \cite{1}
describes impressively well all experimental data
\cite{2}. This description depends on two unknown parameters, the
top mass, $m_t$, and the Higgs mass, $m_H$.
The present experimental accuracy of the electroweak measurements justifies
to look for indirect bounds on these masses.
In this paper we present the $\chi^2$ distributions
resulting from the most precise and recent electroweak data as functions
of $m_t$ and $m_H$. Similar analyses can be found in the literature
\cite{3,4,5}. We pay particular attention to the Higgs penalty
function describing the Higgs production limit. It plays an important
role due to the low sensitivity of the data on $m_H$ and to the fact
that the minimum of the $\chi^2$, if only indirect data are used,
is (at present) well below the direct production limit.
Actually, the present sensitivity of the data on $m_H$ will prove to
be too low to obtain stringent, statistically significant, limits.
However the sensitivity could be greatly improved in the near future,
provided that LEP errors get reduced and the top is discovered.

\section*{Data}

\noindent
The experimental values used in the fits are gathered in Table 1.
The detailed way LEP excludes Higgs production
can be found in Ref. \cite{6}. For our discussion it is
enough to realize that the final result of the search is that no event with
a given signature is found, whereas
some events should be present if the Higgs
boson would have been produced. Of course, the number of expected
events decreases with increasing $m_H$. Therefore, assuming
a Poisson distribution for the number of expected events, values below $m_H$
can be excluded (for instance) at the $95\%\ c\ell$ if the number of
expected events for $m_H$
is larger than or equal to $3$. The number of expected events at LEP,
$N$, as a
function of $m_H$ is shown in Fig.1a (this plot is obtained from the
compilation in Ref. \cite{7}).
The confidence level, $c\ell$, corresponding to seeing no events when $N$
events are expected is equal to $ 1 - e^{-N} $; and
this confidence level is transformed into the corresponding
number of standard deviations, $x$, for a single sided $N(0,1)$ distribution
using the relation
$$ \frac{c\ell + 1}{2} = \frac{1}{\sqrt{2\pi}} \int_{-\infty}^x e^{-t^2} dt. $$
\noindent
In this way $x$ can be written
as a function of $m_H$. This function is plotted in
Fig.1b, and the best linear fit,
$$ x = 15.81 - 0.24 \ m_H,  $$
is superimposed on it. Thus, the corresponding contribution
to $\chi^2$ is given by the penalty function
$$\Delta\chi^2_H =
\left\{ \begin{array}{cl}
 \left(
{\displaystyle \frac{65.0-m_H(GeV)}{4.1}} \right)^2 &
\mbox{ if $m_H \leq 65 \ GeV,$ }$$ \\
0 & \mbox{ if $m_H > 65 \ GeV.$ }
\end{array}
\right.$$

 In the case of the top quark the detailed way TEVATRON excludes top production
can be found in Ref. \cite{8}.
In this case, however, the minimum of the $\chi^2$, is well above the direct
production
limit, and then the
corresponding penalty function plays no important role in the fit.
The strategy for the top search is also
different. What is studied in the top case
is the presence of
an excess of events in some distributions (basically in the transverse mass
distribution)
due to top production.
These distributions are fitted with a linear superposition of the
standard model predictions with no top quark
plus the top production.
The result of this fit
for each top mass is a probability density distribution ${\em L}(\rho)$
for the variable $\rho$ defined as the ratio of the number of fitted to
the number of expected
($\sigma_t(m_t)$) top events for such a mass. No top events corresponds
to $\rho = 0$. To exclude a top mass (for instance) with a $95\% \ c\ell$
we associate to each mass $m_t$ the value of $\rho_0(m_t)$ which defines
for the probability density corresponding to this mass an area,
$\int_0^{\rho_0(m_t)} {\em L}(\rho) d\rho$, equal to the
$95\%$ of the total area or probability,
$\int_0^{\infty} {\em L}(\rho) d\rho$.
Then $\rho_0(m_t)=1$ gives the $m_t$ limit below which the top mass
is excluded.
To make this computation for different confidence levels
we have assumed that
$\rho_0(m_t)\sigma_t(m_t)$ is independent of $m_t$ in the region of
interest. (Note that this is equivalent to assume that the shape of the
top distribution is independent of the top mass in this region.)
To translate the $c\ell$ to which a given top mass is excluded
into a number of standard deviations we proceed as in the case of the
Higgs mass. The corresponding fit gives the approximate penalty function
$$\Delta\chi^2_t =
\left\{ \begin{array}{cl}
 \left(
{\displaystyle \frac{108-m_t(GeV)}{9}} \right)^2 &
\mbox{ if $m_t \leq 108 \ GeV,$ }$$ \\
0 & \mbox{ if $m_t > 108 \ GeV.$ }
\end{array}
\right.$$

 The $\chi^2$ also includes
neutral current data: neutrino-quark data, $\nu q$ \cite{9},
the latest
neutrino-electron results, $\nu_{\mu} e$ \cite{10}, and
the parity violation in atoms data, $e H$ \cite{11,12}; the $W$ mass
limits: $M_W$ \cite{13} and $\frac{M_W}{M_Z}$ \cite{14}; the LEP
results presented at the
Joint International Lepton--Photon Symposium and Europhysics
Conference on High--Energy Physics  in August 1991
\cite{15}
(with the correlation matrix given in Ref. \cite{15p}); and the strong
coupling constant, $\alpha_s$ \cite{16}. We take for $\alpha_s$ the
latest (most precise) ALEPH
measurement using hadronic $Z$ decays. The definite
$\alpha_s$ value is important to fix the $\chi^2$ value at the minimum
but modifies little the relative $\chi^2$ distribution and does not
affect the conclusions.

\section*{Theory}

\noindent
We use the on-shell scheme of Ref. \cite{17}
with the electromagnetic coupling constant, $\alpha$, and the
Fermi constant, $G_{\mu}$, \cite{9}
as input parameters.
For the calculation of the observables in the $Z$ physics sector
two independent electroweak libraries have been used: the one from
G. Burgers and W. Hollik \cite{HOLLIK} and the one from the Dubna-Zeuthen
group \cite{BARDIN}. In spite of the fact that they use a completely different
calculational scheme, their results are in very good agreement
\cite{agreement}. We have upgraded the first one to incorporate:

\begin{itemize}
\item{} The missing relevant pieces of two-loop corrections \cite{two-loop};
\item{} The dominant QCD corrections to $m_t$ dependent terms
         \cite{mt*alfas,QCDbbar};
\item{} The  updated calculation of the QCD corrections to the decay
         $Z \rightarrow b \bar{b}$ \cite{QCDbbar}; and,
\item{} $O(\alpha_s^3)$ corrections to the decay widths \cite{alfas**3}.
\end{itemize}
Similar upgrades have been implemented in the Dubna-Zeuthen
library by their authors. A comparison of the  present versions of both
programs shows an even better agreement than the one previously quoted.
While the numbers and plots shown in this study have been obtained with the
first library, we have explicitly checked that the changes
are negligible if the
Dubna-Zeuthen library is used instead.
In our fits we use the explicit expressions for all the observables,
in particular for the neutral current ones \cite{25}.
We have checked that
reducing the full experimental information for the latter to the
equivalent electroweak mixing angle value, our fits and conclusions
change little.

\section*{Interpretation}

\noindent
Our results are summarized in Figs.2-8. Fig.2a shows the
contribution to the $\chi^2$ distribution
of  neutral current  and
$p\bar{p}$ collider data; Fig.3a that
of LEP data and $\alpha_s$; Fig.4a their sum; and Fig.5a that
of all the data including the penalty functions.
Thus Figs.2a-4a give partial contributions to Fig.5a.
At each point in the $m_t-m_H$ plane we
minimize with respect to $M_Z$ and $\alpha_s$ ($G_{\mu}$ and $\alpha$
being fixed).
Thus, the total number of degrees of freedom is $20-2=18$. (The two
degrees of freedom corresponding to the penalty functions are taken
care by the two parameters, $m_t$ and $m_H$, with respect to which
we do not minimize.)
Figs.2b-5b show the same fits but replacing the
central experimental values of the different observables by their
predicted (standard model) values near the minimum of the $\chi^2$
distribution. In particular we have taken $m_H=70\ GeV$,
$m_t=130\ GeV$. The corresponding standard model predictions
for the observables used in the fits are given in Table 2.
Comparing Figs.2a-4a and Fig.5a we see that the Higgs penalty
function moves the minimum from $m_H=10$ to $70\ GeV$. However,
the main observation to be elaborated below
when comparing Fig.3a and Fig.3b is the apparent change in the LEP
distribution, and then in the global one.
This tells us that some experimental central values manifest a
statistical fluctuation or indicate new physics. This comparison
also proves the actual lack of sensitivity of the data on $m_H$.
To understand this change
it is illuminating to rewrite (in obvious notation)
$$
\begin{array}{rl}
\chi^2(m_H,m_t,...) & ={\displaystyle
 \sum_i \frac{(X_i^{{\rm exp}}-X_i^{{\rm th}}(m_H,m_t,...))^2}
{\sigma_i^2}
= \sum_i \frac{(X_i^{{\rm exp}}-X_i^0+\epsilon_i)^2}
{\sigma_i^2}}  \vspace{.5cm} \\
&   = {\displaystyle
\sum_i \frac{(X_i^{{\rm exp}}-X_i^0)^2-2(X_i^{{\rm exp}}-X_i^0)\epsilon_i
+\epsilon_i^2}
{\sigma_i^2}},
\end{array}
$$
\noindent
where $X_i^0$ are the standard model predictions at the $\chi^2$
minimum and $\epsilon_i$ are the differences between the predicted
values at a given point and the values at the minimum.
The last expression shows that a statistical fluctuation or a relatively
large change in the central value of any observable contributes to
the $\chi^2$ with a relatively large constant term
${\displaystyle \frac{(X_i^{exp}-X_i^0)^2}
{\sigma_i^2}}$ and with a linear term of pronounced slope
${\displaystyle \frac{-2(X_i^{exp}-X_i^0)\epsilon_i}
{\sigma_i^2}}$. However, what gives the sensitivity of the observable
is ${\displaystyle \frac{\epsilon_i^2}{\sigma_i^2}}$.
Comparing Figs.2a-5a and Figs.2b-5b we see that indirect LEP data are
very insensitive to $m_H$ but some observables have a large discrepancy
between their measured central values and the predicted ones.
The global fit to all data (Fig.5a) gives
$$
\begin{array}{rl}
m_H= & {\displaystyle 65_{-4}^{+245}}  \ GeV,\vspace{.5cm}\\
m_t= & {\displaystyle 122_{-20}^{+25}}  \ GeV.
\end{array}
$$
\noindent
Although the fits in Figs.2a-5a
give the actual experimental information, those
in Figs.2b-5b
give the actual sensitivity of the data on $m_H$ and $m_t$.
Hence present indirect electroweak data are not sensitive to $m_H$.
Actually, the present LEP discrepancy can be traced back to the value of
the ratio $R_{\ell}$ and the bottom forward--backward
asymmetry $A_{FB}^{b\overline{b}}$.
This is to say that the data show a higher
sensitivity than expected on $m_H$ due to the pronounced slope
terms (introduced above) resulting from $R_{\ell}$ and
$A_{FB}^{b\overline{b}}$. If this discrepancy is a statistical
fluctuation and it disappears when more data are available,
the subsequent fit will approach Figs.2b-5b. However, it can also happen
that the central values of other observables fluctuate, in which case our
discussion would apply to them. At any rate this will be always the case
(if new physics is discarded) when the preferred $m_H$ value from indirect
LEP data lies below the production limit.
In Figs.6a,b we show the
$\chi^2$ distribution as a function of $m_H$ (minimization with
respect to $m_t$ is understood) for the global fits in Figs.5a,b.
Plotting the same distribution for the different observables
it can be explicitly proven
that only $R_{\ell}$ and $A_{FB}^{b\overline{b}}$ show the
pronounced slope behaviour for the real data, as can be seen in
Fig.6a, whereas the other observables show a small variation, which we
do not plot. Fig.6b confirms that the sensitivity of $R_{\ell}$
and $A_{FB}^{b\overline{b}}$ on $m_H$ is very low, as it is the
sensitivity of the other observables and of the full set.
In fact we plot $\Delta \chi^2$ in Fig.6a, subtracting from each
$\chi^2$ contribution its value at the global minimum:
$10.79$, $0.88$, $2.73$ for the global, $R_{\ell}$,
$A_{FB}^{b\overline{b}}$ contributions, respectively.
Using higher values for $\alpha_s$ would arrange the $R_{\ell}$
deviation from the standard model prediction.
This would translate into a modification of the $m_t$ value at the
minimum in order to conserve the agreement between the measured and the
predicted $\Gamma_Z$ values. However $A_{FB}^{b\overline{b}}$
would still constrain the fit and keep unchanged the value of $m_H$ at the
minimum and the conclusions. Lower values of $\alpha_s$ make worse
the disagreement between the measured and the predicted $R_{\ell}$
values.
It is worth to note that although the top mass limit is significative,
it is strongly correlated to the Higgs mass as can be seen in Fig.5.
For instance if the Higgs mass were known to be $300\pm 30\ GeV$,
the present data would imply $m_t=139^{+21}_{-25}\ GeV$ (see Fig.5a).

\section*{Conclusions}

\noindent
As stressed above, today's data show a fictitious $m_H$ sensitivity due
to a statistical fluctuation or to new physics manifested by the
experimental values of $R_{\ell}$ and $A_{FB}^{b\overline{b}}$
(Figs.2a-6a). As a matter of fact,
the actual sensitivity of the considered
observables on $m_H$ is very low (Figs.2b-6b). The Higgs penalty function
plays an important role in the fits, pushing up the Higgs mass value at
the minimum. What is more interesting is
the expected improvement of the $m_H$ sensitivity in the near future
\cite{26}.
Figs.2-5 show the correlation between $m_H$ and $m_t$ which should
translate, once the top quark is discovered and its mass determined, into
a definite bound on $m_H$. In fact, if present LEP accuracy is improved
as assumed in Table 2, and the top mass is in the preferred range
given above, a meaningful bound on
the Higgs mass (within the standard model framework) could
be obtained. This fact is
illustrated in Figs.7 and 8. In Fig.7 we plot the
$\chi^2$-distribution  using the standard model predictions as
central values and the set of LEP {\it improved}
errors quoted in Table 2. We minimize at each point
with respect to $M_Z$ and
$\alpha_s$. Fig.7 shows that fixing $m_t$
a relatively
strong upper bound on $m_H$ can be deduced.
This is explicitly shown in Fig.8,
where we fix the top mass, $m_t=130\pm 1\ GeV$. In this case
$m_H<315\ GeV$ at 95\% $c\ell$.

\section*{Acknowledgements}

  One of us (M.M.) would like to thank J. Steinberger for triggering the
interest for this study as well as for many suggestions
and A. Blondel and G. Rolandi for discussions.
We have benefitted from discussions with J. Ellis,  with
W. Hollik about the upgrade of his libraries and also
with D. Schaile concerning the actual interpretation of the $\chi^2$
distribution. We acknowledge W. Hollik and the Dubna-Zeuthen group for
making us available their computer programs.

\newpage
\def\journal#1#2#3#4{{\it #1} {\bf #2} (#3) #4}

\newpage

\section*{Table captions}
\begin{description}

\item[Table 1]
Experimental values of the observables used in the fits.

\item[Table 2]
Standard model predictions for the observables used in the fits and
for $m_H=70\ GeV$ and $m_t=130\ GeV$. The {\it improved} error
column gives a guess of future LEP errors.

\end{description}

\vspace{1cm}
\section*{Figure captions}
\begin{description}
\item[Fig.1a{[b]}]
Number of expected events [standard deviations] at LEP as a
function of $m_H$.
\item[Fig.2a{[b]}]
Level contours for the contribution to the $\chi^2$-distribution
of neutral current and $p\bar{p}$ collider data in Table 1
[standard model predictions in Table 2
with the present experimental
errors in Table 1]. Minimization with respect to $M_Z$ and
$\alpha_s$ is understood.
\item[Fig.3a{[b]}]
The same as in Fig.2a[b] but for LEP observables including $\alpha_s$.
\item[Fig.4a{[b]}]
The same as in Fig.2a[b] but for all observables in Table 1.
\item[Fig.5a{[b]}]
The same as in Fig.2a[b] but for all observables in Table 1 plus
the top and Higgs penalty functions given in the text.
\item[Fig.6a{[b]}]
$\chi^2$-distributions as functions of $m_H$ for the global fit
in Fig.5a[b] (solid curve). The
$R_{\ell}$ (dashed curve) and the $A_{FB}^{b\bar{b}}$
(dotted curve) contributions are also shown.
Minimization with respect to $M_Z$, $\alpha_s$
and $m_t$ is understood.
\item[Fig.7]
Level contours for the total $\chi^2$-distribution assuming the
standard model predictions as central values and the set of
improved errors given in Table 2. Minimization with respect to $M_Z$
and $\alpha_s$ is understood.
\item[Fig.8]
$\chi^2$-distribution as a function of $m_H$ for the global
fit in Fig.7 and $m_t=130\pm 1\ GeV$. Minimization with respect
to $M_Z$, $\alpha_s$ and $m_t$ is understood.
\end{description}

\newpage

\begin{center}
\begin{tabular}{||c|c|c|cccc||} \hline\hline
\multicolumn{2}{||c|}{} && \multicolumn{4}{|c||}{} \\
\multicolumn{2}{||c|}{Quantity} & Experimental Value
&\multicolumn{4}{|c||}{Correlation Matrix} \\
\multicolumn{2}{||c|}{} && \multicolumn{4}{|c||}{} \\ \hline\hline
&&&&&& \\
& $g_L^2$ & $0.2977 \pm 0.0042$ & 1.&&& \\
$\nu q$ & $g_R^2$ & $0.0317 \pm 0.0034$ &&1.&& \\
& $\theta_L$ & $2.50 \pm 0.03$ &&&1.&  \\
& $\theta_R$ & $4.59^{+0.44}_{-0.27}$ &&&&1.  \\
\hline
$\nu_{\mu}e$ & $g_A^e$ & $-0.503 \pm 0.018 $ &1.&$-0.05$&& \\
& $g_V^e$ & $-0.025 \pm 0.020$ &&1.&& \\
\hline
& $C_{1u}+C_{1d}$ & $0.144 \pm 0.007$&1.&&&  \\
$eH$ & $C_{1u}-C_{1d}$ & $-0.60 \pm 0.09$ &&1.&&  \\
 & $C_{2u}-C_{2d} $ &$ -0.05 \pm 0.11$ &&&1.& \\
\hline
$p\bar{p}$ & $M_W$ & $79.91 \pm 0.39$ GeV &&&&\\
           &$M_W/M_Z$ &$ 0.8813 \pm 0.0041$ &&&&\\
\hline
& $M_Z$ & $91.175 \pm 0.02$1 GeV & 1. & 0.09 & 0.01 & 0.00 \\
& $\Gamma_Z$ & $2487 \pm 10$ MeV & & 1. & --0.25 & --0.07 \\
  & $\sigma_h^0$ & $41.36 \pm 0.23$ nb & & & 1. & 0.18 \\
& $R_{\ell}$ & $20.92 \pm 0.11$  &&&& 1. \\
\cline{2-7}
&&&&&& \\
LEP& From $A_{FB}$: &&&&& \\
 & ${\displaystyle \left(\frac{g_V^{\ell}}{g_A^{\ell}}(M_Z)\right)^2}$
& $0.0048\pm 0.0012$ & 1.&&& \\
&&&&&& \\
 & From $A^{{\rm pol}}(\tau)$: &&&&& \\
 & ${\displaystyle \frac{g_V^{\ell}}{g_A^{\ell}}(M_Z)}$
& $0.072\pm 0.017$ && 1.&& \\
&&&&&& \\
& From $b\bar{b}$-asymmetry: &&&&& \\
& $A_{FB}^{b\bar{b}}(M_Z)$ & $0.132\pm 0.022$ &&& 1. & \\
&&&&&& \\
 & From $q\bar{q}$-asymmetry: &&&&& \\
& $\sin^2\overline{\theta}_W(M_Z)$ & $0.2303\pm 0.0035$ &&&& 1.\\
\cline{2-7}
&&&&&& \\
& $\alpha_s$ & $0.125\pm 0.005$ &&&& \\
&&&&& & \\ \hline\hline
\end{tabular}
\end{center}
\begin{center}
Table 1
\end{center}

\newpage

\begin{center}
\begin{tabular}{||c|c|c|c||} \hline\hline
\multicolumn{2}{||c|}{} & & \\
\multicolumn{2}{||c|}{Quantity} & Standard model prediction
&{\it Improved} errors \\
\multicolumn{2}{||c|}{} & & \\ \hline\hline
&&& \\
& $g_L^2$ & 0.3012 & 0.0042 \\
$\nu q$ & $g_R^2$ & 0.0302 & 0.0034 \\
& $\theta_L$ & 2.46 & 0.03  \\
& $\theta_R$ & 5.18 &0.44 \\
\hline
$\nu_{\mu}e$ & $g_A^e$ & $-0.505$ & 0.018  \\
& $g_V^e$ & $-0.048$ & 0.020 \\
\hline
& $C_{1u}+C_{1d}$ & 0.146 & 0.007  \\
$eH$ & $C_{1u}-C_{1d}$ & $-0.54$ & 0.09 \\
 & $C_{2u}-C_{2d} $ & $-0.09$& 0.11 \\
\hline
$p\bar{p}$ & $M_W$ & 80.14 GeV & 0.06 GeV\\
           &$M_W/M_Z$ & 0.8790 &0.0041\\
\hline
& $M_Z$ & 91.175 GeV & 0.005 GeV  \\
& $\Gamma_Z$ & 2489 MeV & 4 MeV \\
  & $\sigma_h^0$ & 41.41 nb & 0.08 nb  \\
& $R_{\ell}$ & 20.81 & 0.02  \\
\cline{2-4}
&&& \\
LEP& From $A_{FB}$: && \\
 & ${\displaystyle \left(\frac{g_V^{\ell}}{g_A^{\ell}}(M_Z)\right)^2}$
& 0.0048 & 0.0006 \\
&&& \\
 & From $A^{{\rm pol}}(\tau)$: && \\
 & ${\displaystyle \frac{g_V^{\ell}}{g_A^{\ell}}(M_Z)}$
& 0.069 & 0.009 \\
&&& \\
& From $b\bar{b}$-asymmetry: && \\
& $A_{FB}^{b\bar{b}}(M_Z)$ & 0.097 & 0.010 \\
&&& \\
 & From $q\bar{q}$-asymmetry: && \\
& $\sin^2\overline{\theta}_W(M_Z)$ & 0.2320 & 0.0035\\
\cline{2-4}
&&& \\
& $\alpha_s$ & 0.125 & 0.005 \\
&& & \\ \hline\hline
\end{tabular}
\end{center}
\begin{center}
Table 2
\end{center}

\end{document}